\documentclass
[superscriptaddress,secnumarabic,amssymb,amsmath,nobibnotes,aps,prd,showkeys,showpacs,twocolumn,nofootinbib]{revtex4}%
\usepackage{bm}
\usepackage{hyperref}
\usepackage{mathrsfs}
\usepackage{xcolor,color,graphicx,graphics}
\usepackage[all]{xy}
\usepackage{latexsym,amssymb,amsmath,amsfonts} %\usepackage{amstex}
\usepackage[english]{babel} %, spanish, portuguese
\usepackage[OT1]{fontenc}
\usepackage{makeidx}
\usepackage{hyperref}
\usepackage{color,graphicx,graphics,wrapfig,epsf}%,psfig
\usepackage{subfig}
\usepackage{mwe}
\usepackage{orcidlink} 
\hypersetup{urlcolor=BlueViolet,
	    citecolor=Plum,
	    linkcolor=PineGreen}

\usepackage[labelfont=bf]{caption}
\begin{document}

%=================================================================
% Full title of the paper (Capitalized)
\title{Bose and Fermi gases in metric-affine gravity and linear Generalized Uncertainty Principle}

\author{Aneta Wojnar\orcidlink{0000-0002-1545-1483}}
\thanks{Corresponding author}
\email[E-mail: ]{awojnar@ucm.es}
\affiliation{Department of Theoretical Physics \& IPARCOS, Complutense University of Madrid, E-28040, 
Madrid, Spain}

\author{D{\'e}bora Aguiar Gomes\orcidlink{0000-0003-0079-7323}}
\email[E-mail: ]{debora.aguiar.gomes@ut.ee}
\affiliation{Laboratory of Theoretical Physics, Institute of Physics, University of Tartu, W. Ostwaldi 1, 50411 Tartu, Estonia}

\begin{abstract}
We examine the relationship between Palatini-like theories of gravity and models incorporating linear generalized uncertainty principles. Additionally, we delve into the thermodynamics of systems comprising both Bose and Fermi gases. Our analysis encompasses the equations of state for various systems, including general Fermi gases, degenerate Fermi gases, Boltzmann gases, Bose gases such as phonons and photons, as well as Bose-Einstein condensates and liquid helium.
\end{abstract}

\maketitle

%keywords: 
%Bose gas, Fermi gas, modified gravity, liquid helium, Bose-Einstein condensate, finite temperature corrections, linear GUP

\section{Introduction}

The consistency of equations within Modified Gravity (MG) is brought into question by numerous indications from prior investigations. For instance, the dependence of chemical potential on gravity suggests that modifications in the description of gravitational fields would have an impact \cite{kulikov1995low}. 
Modified Gravity has been demonstrated to alter the geodesic deviation equation on the surface of stars, resembling Hook's law and introducing corrections to the polytropic equation of state \cite{kim2014physics}. Moreover, microscopic quantities such as opacity exhibit modifications, implying a need for an effective treatment \cite{sakstein2015testing}. Thermodynamic laws, stellar stability criteria, and properties of Fermi gases also display corrections stemming from gravitational proposals \cite{Wojnar:2016bzk,Wojnar:2017tmy,Sarmah:2021ule,wojnar2023fermi}.

Theoretical descriptions of thermonuclear processes within stars' interiors undergo changes under modified gravity, thereby affecting computations of energy generation rates \cite{sakstein2015hydrogen,Olmo:2019qsj,crisostomi2019vainshtein,rosyadi2019brown,wojnar2021lithium}. Some gravity theories introduce a dependence of elementary particle interactions on local energy-momentum distributions \cite{delhom2018observable}. Additionally, specific heats, Debye temperatures, and crystallization processes in white dwarfs are influenced by the gravity model \cite{Kalita:2022trq,Wojnar:2024kzy}. Furthermore, chemical reaction rates, which are influenced by gravity \cite{lecca2021effects}, are expected to change with modifications to this interaction.

Relativistic effects in equations of state, when neglected in Tolman-Oppenheimer-Volkoff equations derived from General Relativity (GR), result in an underestimation of compact star limiting masses. Equations of state in curved spacetime for degenerate stars explicitly depend on metric components, leading to alterations in chemical potentials and temperatures \cite{hossain2021equation,hossain2021higher,li2022we,chavanis2004statistical}. Moreover, thermodynamic quantities and equations of state are further modified when (pseudo-)scalar fields, such as axions, are considered \cite{sakstein2022axion}. 

The emerging link between MG and Generalized Uncertainty Principle (GUP), as shown in \cite{Wojnar:2023bvv,Ali:2024tbd}, opens avenues for testing gravitational theories within Earth-based laboratories. Additionally, techniques advanced by either community can be utilized to evaluate proposals concerning MG or GUP interchangeably \cite{Wojnar:2024xdy}. Notice that such a correspondence is also expected; it is believed that MG is an effective theory of \textit{some} Quantum Gravity. Many proposals emphasize integrating the quantum structure of space-time and deforming associated quantum phase spaces, leading to the generalization of the Heisenberg uncertainty principle. This emphasis is underscored for the potential measurable effects it offers \cite{Pachol:2023bkv,Pachol:2023tqa,Kozak:2023vlj}. GUP has proven valuable in predicting quantum gravity effects, as evident in various models found in \cite{kempf1995hilbert,maggiore1993generalized,maggiore1994quantum,chang2002exact,chang2002effect,moussa2015effect}. Despite variations in mathematical structures, a shared characteristic among most of these models is the existence of a minimum length scale, anticipated to be approximately the Planck length, $L_P\sim \sqrt{\frac{\hbar G}{c^3}}$, as detailed in \cite{bishop2020modified,bishop2022subtle,segreto2023extended}.

The GUP models introduce modifications to equations of state and microscopic variables, stemming from the interplay between special relativity and gravity. This proposition incorporates a dispersion relation involving energy, mass, and momentum within Heisenberg's Uncertainty Principle, incorporating constants such as the speed of light and the gravitational constant \cite{moussa2015effect,rashidi2016generalized,belfaqih2021white,mathew2021existence,hamil2021new,gregoris2022chadrasekhar}.

Below, we will delve into a review of previous findings regarding ideal gases, derived from the connection between Ricci-based gravity and linear GUP. Before proceeding, we will provide a brief recap of this correspondence. Additionally, we introduce, for the first time, a theoretical framework for ideal and Boltzmann gases, incorporating Fermi statistics in both idealized and realistic scenarios, along with a simplified portrayal of Bose gases, such as photon and phonon gases, within the context of Ricci-based and linear GUP models.

%%%%%%%%%%%%%%%%%%%%%%%%%%%%%%%%%%%%%%%%%%
\section{Metric-affine gravity and linear GUP correspondence}

The class of metric-affine gravity theories under consideration is defined by the action:

\begin{equation} \label{eq:actionRBG}
\mathcal{S}=\int d^4 x \sqrt{-g} \mathcal{L}_G(g_{\mu\nu},R_{\mu\nu}) +
\mathcal{S}_m(g_{\mu\nu},\psi_m) \ .
\end{equation}
Here, $g$ denotes the determinant of the space-time metric $g_{\mu \nu}$, and $R_{\mu \nu}$ represents the symmetric Ricci tensor, which relies solely on the affine connection $\Gamma \equiv \Gamma_{\mu\nu}^{\lambda}$. To construct the gravitational Lagrangian $\mathcal{L}_G$ as a scalar function using powers of traces of ${M^ \mu}{\nu} \equiv g^{\mu \alpha}R_{\alpha\nu}$, the object ${M^ \mu}_{\nu}$ is introduced.

The matter action is described by:

\begin{equation}
\mathcal{S}_m=\int d^4 x \sqrt{-g} \mathcal{L}_m(g_{\mu\nu},\psi_m).
\end{equation}
In this framework, the matter action couples minimally to the metric, neglecting the torsion (the antisymmetric part of the connection), akin to minimally coupled bosonic fields. This simplification extends to fermionic particles, such as degenerate matter, effectively described by a fluid approach represented by the perfect fluid energy-momentum tensor \cite{alfonso2017trivial}. Focusing on the symmetric part of the Ricci tensor avoids potential ghostlike instabilities \cite{Borowiec:1996kg,Allemandi:2004wn,beltran2019ghosts,jimenez2020instabilities}. This framework encompasses various gravity theories, including GR, Palatini $f(R)$ gravity, Eddington-inspired Born-Infeld gravity \cite{vollick2004palatini}, and its extensions \cite{jimenez2018born}.

The gravitational action encapsulates theories that, despite complex field equations, can be conveniently reformulated \cite{jimenez2018born}:

\begin{equation} \label{eq:feRBG}
{G^\mu}_{\nu}(q)=\frac{\kappa}{\vert \hat{\Omega} \vert^{1/2}} \left({T^\mu}_{\nu}-\delta^\mu_\nu \left(\mathcal{L}_G + \frac{T}{2} \right) \right) \ .
\end{equation}

Here, $\vert\hat{\Omega}\vert$ denotes the determinant of the deformation matrix, and $T$ represents the trace of the energy-momentum tensor of matter fields. The Einstein tensor ${G^\mu}_{\nu}(q)$ is linked to a tensor $q_{\mu\nu}$, where the connection $\Gamma$ adopts the Levi-Civita connection of $q_{\mu\nu}$:

\begin{equation}
\nabla_{\mu}^{\Gamma}(\sqrt{-q} q^{\alpha \beta})=0.
\end{equation}

In this formalism, the tensor $q_{\mu\nu}$ is related to the space-time metric $g_{\mu\nu}$ through:

\begin{equation}\label{eq:defmat}
q_{\mu\nu}=g_{\mu\alpha}{\Omega^\alpha}_{\nu} ,
\end{equation}
where the deformation matrix ${\Omega^\alpha}_{\nu}$ is theory-dependent and determined by the gravitational Lagrangian $\mathcal{L}_G$. These theories yield second-order field equations, reducing to GR counterparts in vacuum (${T_\mu}^{\nu}=0$), implying no extra degrees of freedom propagate in these theories beyond the usual two polarizations of the gravitational field.

The nonrelativistic limit of the field equations \eqref{eq:feRBG} is particularly intriguing. In Palatini $f(R)$ \cite{Toniato:2019rrd} and EiBI \cite{banados2010eddington,pani2011compact} gravities, the Poisson equation assumes the form:

\begin{equation}\label{poisson}
\nabla^2\phi = \frac{\kappa}{2}\Big(\rho+\bar\alpha\nabla^2\rho\Big).
\end{equation}
Here, $\phi$ denotes the gravitational potential, $\kappa=8\pi G$, and $\bar\alpha$ represents a theory parameter. In Palatini $f(R)$, $\bar\alpha=2\bar\beta$, with $\bar\beta$ accompanying the quadratic term, while in EiBI, $\bar\alpha=\epsilon/2$, where $\epsilon=1/M_{BI}$ and $M_{BI}$ denotes the Born-Infeld mass. The resemblance in the Poisson equation between these two gravity proposals is not coincidental; the EiBI gravity, in the first-order approximation, reduces to Palatini gravity with the quadratic term \cite{pani2012surface}. Furthermore, only the quadratic term $R^2$ influences the non-relativistic equations, as higher curvature scalar terms enter the equations at the sixth order \cite{Toniato:2019rrd}.

As indicated in \cite{Wojnar:2023bvv}, the augmentation in the Poisson equation \eqref{poisson} can be interpreted as a modification to the Fermi gas at finite temperature. This modification arises when considering a deformation of the phase space represented by the integral:

\begin{equation}\label{sumint}
\frac{1}{(2\pi \hbar)^3} \int \frac{d^3xd^3p}{(1-\sigma p)^{d}},
\end{equation}
where $d=1$ corresponds to Palatini-like gravity theories. The connection between the deformation parameter $\sigma$ and the Palatini parameter $\bar\beta$ is established as:
\begin{equation}\label{palparam}
\sigma = \frac{4\pi G}{K_2}\bar\beta\,\,\,\,\text{and}\;\;\; K_2 = \frac{3}{\pi} \frac{h^3N_A^2}{m_e \mu_e^2},
\end{equation}
with $m_e$ being the electron mass, $\mu_e$ the mean molecular weight per electron, and other constants holding their usual definitions.

This correspondence enables the expression of a general partition function in three dimensions within a large volume
\begin{equation}\label{partition}
\mathrm{ln}Z = \frac{V}{(2\pi \hbar)^3}\frac{g}{a}\int \mathrm{ln}\left[1+az e^{-E/k_BT}\right] \frac{d^3p}{(1-\sigma p)^{d}} \ ,
\end{equation}
where $V:=\int d^3x$ denotes the volume of the cell in configuration space, and setting $a=1$ ($a=-1$) corresponds to a system of fermionic (bosonic) particles with energy states $E_p$. The fugacity is represented by $z=e^{\mu/k_BT}$, with $\mu$ as the chemical potential, and $g$ indicating the spin of a particle.
    
Similar to the GUP incorporating linear $p$-modifications \cite{cortes2020deformed,ali2009discreteness,ali2011minimal,abac2021modified,vagenas2019gup,tawfik2014generalized}, our approach integrates a deformed phase space measure characterized by the parameter $\sigma$. In the context of GUP, the deformation is derived through the utilization of the Liouville theorem \cite{vagenas2019linear}. Consequently, the effective $\hbar$ depends on the momentum $p$ in the generalized uncertainty relation, leading to a momentum-dependent size of the unit cell for each quantum state in phase space.

With this modified partition function, one can straightforwardly derive thermodynamic variables for the requisite statistics, focusing primarily on pressure, number of particles, internal energy, and specific heat, which are respectively given by:

\begin{align}
    P=&\,k_B T\frac{\partial}{\partial V} \mathrm{ln}Z, \label{therm1}\\
    n=&\,k_B T\frac{\partial}{\partial \mu} \mathrm{ln}Z\mid_{T,V}, \label{therm2}\\
    U=&\,k_B T^2\frac{\partial}{\partial T} \mathrm{ln}Z\mid_{z,V} \label{therm3}\\
    C_V=\,&\frac{\partial U}{\partial T}\mid_{V} \label{therm4}.
\end{align}

%%%%%%%%%%%%%%%%%%%%%%%%%%%%%%%%%%%%%%%%%%
\section{Boltzmann gas}

In the case of high temperature, thermodynamic systems described by Bose or Fermi statistics behave as the Boltzmann one. 
The Maxwell-Boltzmann statistics corresponds to the limit
$a \rightarrow 0$,
from which the general partition function \eqref{partition} acquires the form:
\begin{equation}\label{MBpartition}
\mathrm{ln}Z = \frac{V}{(2\pi \hbar)^3}\int z e^{-E/k_BT}\frac{d^3p}{(1-\sigma p)^{d}} \ .
\end{equation}
Moreover, the distribution function for this statistics is given by
\begin{equation}
    f_{\text{MB}}(E) = ze^{-E/k_{\text{B}}T},
\end{equation}
corresponding to the regime where Fermi-Dirac, Bose-Einstein and Maxwell-Boltzmann statistics become identical, i.e., $(E-\mu)/k_{\text{B}}T >>1$.

Taking the partition function \eqref{MBpartition}, we can easily obtain
\begin{equation}
    P = \frac{1}{2 \pi^2 \bar{h}^3} \int \frac{1}{3}p^3\, _{2}F_{1}\left(3,d,4,p\sigma\right) f_{\text{MB}}(E) \frac{c^2 p}{E}dp,
\end{equation}
\begin{equation}
n= \frac{V}{2 \pi^2 \hbar^3}\int f_{\text{MB}}(E) \frac{ p^2 dp}{(1-\sigma p)^{d}}
\end{equation}
\begin{equation}
U= \frac{V}{2 \pi^2 \hbar^3}\int Ef_{\text{MB}}(E) \frac{ p^2 dp}{(1-\sigma p)^{d}}.
\end{equation}

\section{Fermi statistics}

In this extended scenario, utilizing (\ref{partition}) and (\ref{therm1}) with $a=1$ for fermions and $g=2$ for electrons, we derive the microphysical description of the system employing the Fermi-Dirac distribution $f(E)$:
\begin{equation}
f(E)=\left(1+z^{-1} e^{E/k_BT}\right)^{-1},
\end{equation}
resulting in the expression for pressure as:
\begin{equation}\label{pressuregen}
P= \frac{1}{\pi^2 \hbar^3}\int \frac{1}{3}p^3\, _{2}F_{1}\left(3,d,4,p\sigma\right) f(E) \frac{c^2 p}{E}dp,
\end{equation}
where $_{2}F_{1}$ represents the hypergeometric function. Meanwhile, the particle number density and internal energy are given by:
\begin{equation}
n= \frac{V}{\pi^2 \hbar^3}\int f(E) \frac{ p^2 dp}{(1-\sigma p)^{d}}
\end{equation}
\begin{equation}
U= \frac{V}{\pi^2 \hbar^3}\int Ef(E) \frac{ p^2 dp}{(1-\sigma p)^{d}}.
\end{equation}
For the case when $|\sigma p| < 1$, the general form of the pressure \eqref{pressuregen} can be written as series 
\begin{equation}
      P=  \frac{1}{\pi^2 \hbar^3}\int \frac{p^3}{3} \left( 
      \sum_{k=0}^{\infty} \frac{ (d)_k\left(\frac{3}{2}\right)_k(\sigma p)^k}{ \left(\frac{5}{2}\right)_k k!}
      \right) f(E) \frac{c^2 p}{E}dp,
\end{equation}
while taking into account only the first two terms of the series (as we consider the gravity deformation only up to linear terms in $\sigma$), we have
\begin{equation}\label{pressure}
P= \frac{1}{\pi^2 \hbar^3}\int \frac{1}{3}p^3 \left(1+\frac{3d}{4}\sigma p\right) f(E) \frac{c^2 p}{E}dp.
\end{equation}
We will now discuss its form in more detail.

\subsection{Non-relativistic and relativistic degenerate Fermi gas}\label{deg}

We will now shift our focus to a specific form of matter that holds significant importance in stellar physics: degenerate gases. These gases are utilized to characterize the dense cores of stellar and substellar objects, as well as degenerate matter in compact stars. In the toy model that we are employing here, all states with energies lower than the Fermi energy level are occupied, while all states with higher energies remain empty at absolute zero temperature ($T\rightarrow0$). Consequently, the chemical potential $\mu$ takes on the value of the Fermi energy $E_F$. In such scenarios, the Fermi-Dirac distribution takes the form of a step function:
\[ f(E)=  \left\{
\begin{array}{ll}
      1 & \mbox{if } E\leq E_F \\
      0 & \mbox{otherwise.} \\
\end{array} 
\right. \]
Consequently, in equation (\ref{pressure}), integration is carried out up to the Fermi energy $E_F$:
\begin{equation}
    P = \frac{1}{\pi^2 \hbar^3} \left[ \frac{(2m_e E_F)^{5/2}}{6 m_e} + \frac{\sigma d}{3}\frac{(2m_e E_F)^3}{8m_e}\right], 
\end{equation}
which corresponds to the effective pressure in Palatini theories when $d=1$ (see \cite{Wojnar:2023bvv}). Using the definition of the electron degeneracy
\begin{equation}\label{degparameter}
    \psi = (\beta E_F)^{-1} = \frac{k_{\text{B}}T}{A}\left( \frac{\mu_e}{\rho} \right)^{2/3}
\end{equation}
we can rewrite it as a function of the density such that one deals with
 the non-relativistic polytrope EoS.
Let us write it in a more convenient form for the further analysis:
\begin{equation}
P = K \rho^\frac{5}{3} + \sigma K_2 \rho^2,
\end{equation}
where $K_2$ is given by \eqref{palparam} while 
$$K=\frac{1}{20}\left(\frac{3}{\pi}\right)^\frac{2}{3}\frac{h^2}{m_e(\mu_e m_u)^\frac{5}{3}}$$ 
refers to the standard polytropic parameter for $n=3/2$.

On the other hand, the relativistic polytropic EoS is given simply by inserting $E=pc$ in \eqref{pressure} and using the relation between the pressure and density (recall that for electrons $g=2$)
\begin{equation}
    p=\left(\frac{6\pi^2\hbar^3\rho}{gm_e} \right)^\frac{1}{3},
\end{equation}
providing
\begin{equation}
    P = \frac{\hbar c}{4} \left( \frac{3\pi^2}{m_e^4}\right)^{1/3} \rho^{4/3} + \frac{9}{20} d\sigma \hbar^2 c \left(\frac{3^2 \pi^4}{m_e^5}\right)^{1/3
} \rho^{5/3}.
\end{equation}
Note that the phase-space deformation provides a mixture of two polytropes; that is, in the case of non-relativistic case, the modification is in the form of the rigid polytrope while relativistic one has a term resembling the non-relativistic polytrope ($n=3/2$).

\section{Bose statistics}
Let us briefly recall the basic equations describing a system of $N$ spinless, non-interacting particles in Palatini gravity \cite{Wojnar:2024xdy}. The Hamiltonian is given by $ H= \sum^N_{i=1} \frac{p_i^2}{2m}$, with $p_i^2=\mathrm{p}_i \cdot \mathrm{p}_i$ and $\mathrm{p}_i$ being the momentum operator of the single-particle with energy $E_p=p^2/2m$. Since the grand partition function of an ideal Bose gas for such a system is given by ($\beta =:(k_BT)^{-1}$)
\begin{equation}
    Z = \prod_p \frac{1}{1-z e^{-\beta E_p}},
\end{equation}
an average occupation number for a state $p$ is (the total number of particles $N=\sum_{p}   \langle n_p \rangle$)
\begin{equation}
    \langle n_p \rangle = -\frac{1}{\beta} \frac{\partial}{\partial E_p}\mathrm{ln}Z =\frac{ze^{-\beta E_p}}{1-ze^{-\beta E_p}},
\end{equation}
while the equation of state 
\begin{equation}\label{p1}
    \beta{PV}=-\sum_p \mathrm{ln}(1-ze^{-\beta E_p}).
\end{equation}
Considering $V\rightarrow\infty$ and defining the specific volume $v=V/N$, the above series can be written as
\begin{align}
   \beta P =& - \frac{4\pi}{(2\pi \hbar)^3} \int_0^\infty\frac{dp p^2}{1-\sigma p}
    \mathrm{ln}\left[1-z e^{-\beta\frac{p}{2m}}\right]- \frac{\mathrm{ln}(1-z)}{V} ,\\
    \frac{1}{v} =&  \frac{4\pi}{(2\pi \hbar)^3} \int_0^\infty\frac{dp p^2}{1-\sigma p}
    \frac{1}{z^{-1} e^{\beta\frac{p}{2m}}-1} + \frac{1}{V} \frac{z}{1-z}.
\end{align}
Taking into account only the terms linear in the parameter $|\alpha|=: |\sigma|\sqrt{2mk_BT}$, one writes\footnote{To see how the series converge, see \cite{Wojnar:2024xdy}.}
\begin{align}
   \beta P 
   &= \frac{1}{\lambda^3}\left[ g_{5/2}(z) + \frac{2\alpha }{\pi}\mathrm{Li}_3(z)\right] - \frac{\mathrm{ln}(1-z)}{V} , \label{Eos}\\
    \frac{1}{v} &= \frac{1}{\lambda^3} \left[ g_{3/2}(z) + \frac{2\alpha }{\pi} \mathrm{Li}_2(z) \right] + \frac{1}{V} \frac{z}{1-z}, \label{v}
\end{align}
where $\lambda = \sqrt{\frac{2\pi\hbar^2}{mk_BT}}$ is the thermal wavelength, $g_{m}(z)=\sum_{n=1}^\infty \frac{z^n}{n^{m}}$ and $\mathrm{Li}_n(z)$ is the polylogarithm function which can be expressed for $|z|<1$ as $ \mathrm{Li}_n(z) = \sum_{k=1}^{\infty} \frac{z^k}{k^n}  .$
The internal energy is given simply by
\begin{equation}
   \frac{U}{V}  = -\frac{1}{V} \frac{\partial}{\partial \beta} \mathrm{ln} Z = 
    \frac{3k_BT}{2\lambda^3}\left[ g_{5/2}(z) + \frac{2\alpha }{\pi}\mathrm{Li}_3(z)\right] .
\end{equation}
Comparing it with \eqref{Eos} with the assumption that its last term can be neglected, we deal with a well known relation between the internal energy and temperature $ U=\frac{3}{2}PV$. Notice that in order to study the further properties of ideal Bose gas, one needs to know the fugacity $z$ dependence on the temperature and specific volume $v$. We will come back to this issue in the subsection \ref{conden}.

\subsection{Photon gas}
Since photons are massless, their energy will be simply given by $E=cp$. Its wave number and frequency are related by $p=\hbar k=\hbar \omega/c$ and there are two propagating modes, which are taken into account by multiplying the partition function \eqref{partition} by a factor of 2. The internal energy will be given by
\begin{equation}
    U = \frac{V}{\pi^2} \left[ \frac{\pi^4 (k_{\text{B}}T)^4}{15(\hbar c)^3} + 24 \sigma d \frac{(k_{\text{B}}T)^5 \zeta(5)}{c^4 \hbar^3} \right],
\end{equation}
where the first term corresponds to Stefan's law. Accordingly, the specific heat will be given by
\begin{equation}
    C_{\text{V}} = \frac{V}{\pi^2} \left[ \frac{4 \pi^4 k_{\text{B}}^4T^3}{15(\hbar c)^3} + 120 \sigma d \frac{k_{\text{B}}^5T^4 \zeta(5)}{c^4 \hbar^3} \right]
\end{equation}
while pressure is simply given by $PV=U/3.$

\subsection{Phonon gas}
In spite of not being particles, the vibrations in lattices and in particular its normal modes, phonons, can be mathematically described as bosons. This treatment is specially relevant when considering a solid subject to low temperatures, where phonon interactions are negligible.
In this case, we have three propagation modes, in contrast with photons which only have two, and the allowed frequencies are bounded. The energy per atom is
\begin{equation} \label{energyphonon}
    \frac{U}{N} = 3 \kappa_{\text{B}}T D_3(\beta \hbar \omega) + \frac{9}{4c} \kappa_{\text{B}}T d \sigma \hbar \omega_m D_4(\beta \hbar \omega), 
\end{equation}
where $\omega_m = c \left(\frac{6 \pi^2 N}{V} \right)^{1/3}$ is the maximum frequency and
\begin{equation}
    D_n(x) = \frac{n}{x^n} \int_{0}^{x} \frac{t^n}{e^t-1} dt
\end{equation}
is the Debye function. By defining the Debye temperature via $\kappa_{\text{B}}T_D = \hbar \omega_m$, and using the solutions for 
the Debye functions, we can rewrite \eqref{energyphonon} as 
\begin{equation}
    \frac{U}{N} = 3 \kappa_{\text{B}}T \left( 1-\frac{3}{8} \frac{T_D}{T} \right) + \frac{9}{4c} \kappa_{\text{B}}^2 T_D T d \sigma \left( 1-\frac{2}{5} \frac{T_D}{T} \right) 
\end{equation}
for $T>>T_D$ and
\begin{equation}
    \frac{U}{N} = 3 \kappa_{\text{B}}T \left[\frac{\pi^4}{5} \left( \frac{T}{T_D}\right)^3 \right] + 216 \zeta(5)  \kappa_{\text{B}}T d \sigma (\kappa_{\text{B}}T_D) \left( \frac{T}{T_D}\right)^4
\end{equation}
for $T<<T_D$.

The specific heat, for high and low temperatures, will be
\begin{equation} \label{specificheat}
    \frac{C_\text{V}}{ \kappa_{\text{B}}N} = 3\left[1-\frac{1}{20}\left( \frac{T_D}{T} \right)^2\right] + \frac{9}{4c}d \sigma \kappa_{\text{B}} T_D \left[1-\frac{1}{18} \left( \frac{T_D}{T}\right)^2 \right], \, T>>T_D
\end{equation}
and
\begin{equation}
    \frac{C_\text{V}}{ \kappa_{\text{B}}N} = \frac{12 \pi^4}{5} \left(\frac{T}{T_D}\right)^3 + 1080 \zeta(5)  \frac{d \sigma}{c} \kappa_{\text{B}} T_D  \left(\frac{T}{T_D}\right)^4, \, T<<T_D.
\end{equation}
The first term in \eqref{specificheat} corresponds to the Dulong-Petit law, namely $C_v = 3 N \kappa_{\text{B}}$. However, it should be emphasized that although $T>>T_D$, the temperatures considered here are low enough so that we can consider the phonons to be non-interactive. At very high temperatures, the phonons interaction cannot be overlooked and the current approach will no longer be valid.

\subsection{Bose-Einstein condensate}\label{conden}
The behaviour of the fugacity $z$ depends on the properties of the the functions $g_{3/2}(z)$ and $\mathrm{Li}_2(z)$. Analyzing the equation \eqref{v}, one can derive a modified condition which the ideal Bose gas has to satisfy to become the Bose-Einstein condensate ( $g_{3/2}(1) \approx2.612)$):
\begin{align}
       \frac{\lambda^3}{v} - \frac{2\alpha }{\pi} \zeta(2)=
    g_{3/2}(1).
\end{align}
It clearly provides the critical value for the specific volume (or critical density $n_c=1/v_{cr}$):
\begin{equation}\label{ncrit}
       n_c= \left( \frac{1}{4\pi\hbar^2} \right)^\frac{3}{2} \left[ 
    \zeta(3/2) (2m k_B T)^\frac{3}{2} + \sigma\frac{\pi}{3} (2m k_B T)^2
    \right].
\end{equation}
Moreover, the fugacity dependence on the temperature $T$ and specific volume $v$ is
\[
 z = 
  \begin{cases} 
     1   \mbox{ for }\;\;\;\;\;\;\;\;\;  \frac{\lambda^3}{v} -\frac{\pi\alpha }{3} \geq g_{3/2}(1)& \\
    \mbox{solution of }   \frac{\lambda^3}{v} =
       \left[ g_{3/2}(z) + \frac{2\alpha }{\pi}\mathrm{Li}_2(z) \right]  & \mbox{otherwise. }
  \end{cases}
\]
Hence, the fugacity remains fixed at 1 throughout the Einstein-Bose condensate, indicating a zero chemical potential. In other words, within the region where $\frac{\lambda^3}{v} -\frac{\pi\alpha }{3} \leq g_{3/2}(1)$, we are dealing with the gas phase.

The equation of state and other thermodynamic functions in both regions can be also derived and have the following forms:
\[
 \beta P =
  \begin{cases} 
  \frac{1}{\lambda^3}\left[ g_{5/2}(z) + \frac{2\alpha }{\pi}\mathrm{Li}_3(z)\right]  & \mbox{if } v> v_{cr},  \\
   \frac{1}{\lambda^3}\left[ g_{5/2}(1) + \frac{2\alpha }{\pi}\zeta(3) \right]  & \mbox{if } v< v_{cr},
  \end{cases}
\]

\[
 \frac{U}{N} =\frac{3}{2}Pv=
  \begin{cases} 
\frac{3}{2}  \frac{k_BTv}{\lambda^3}\left[ g_{5/2}(z) + \frac{2\alpha }{\pi}\mathrm{Li}_3(z)\right]  & \mbox{if } v> v_{cr},  \\
 \frac{3}{2}  \frac{k_BTv}{\lambda^3}\left[ g_{5/2}(1) + \frac{2\alpha }{\pi}\zeta(3) \right]  & \mbox{if } v< v_{cr},
  \end{cases}
\]

\[
 \frac{C_V}{Nk_B} =
  \begin{cases} 
\frac{15}{4}\frac{v}{\lambda^3} h_1(T) + \frac{3}{2}\frac{Tv}{\lambda^3} 
h_2(T) \frac{dz}{dT}
  & \mbox{if } v> v_{cr},  \\
\frac{15}{4}\frac{v}{\lambda^3}g_{5/2}(1)+\sigma f_1(T)  & \mbox{if } v< v_{cr},
  \end{cases}
\]
where the functions $h_1\,,h_2,\,f_1$ were defined in \cite{Wojnar:2024xdy}. The vapor pressure takes the following form 
\begin{equation}\label{vapor}
    P_0(T) =   \frac{k_BT}{\lambda^3}\left[ g_{5/2}(1) + \frac{2\sigma\sqrt{2mk_BT} }{\pi}\zeta(3) \right],
\end{equation}
while a modified latent heat resulting from the Clapeyron equations derived from \eqref{vapor} is\footnote{The function $f_2$ is given by \cite{Wojnar:2024xdy}
\begin{equation}
    f_2(T) = \frac{48\sqrt{\frac{\hbar^2\pi}{k_B}}}{5g_{5/2}(1)(g_{3/2}(1)v_{cr})^{1/3}}\zeta(3)
    -\frac{\sqrt{8k_B m T}\zeta(2)}{g_{3/2}(1)} .
\end{equation}}
\begin{equation}\label{latent}
    L=\frac{5}{2} \frac{k_B T g_{5/2}(1)}{g_{3/2}(1)} \left( 1+\frac{\sigma}{\pi} f_2(T)\right).
\end{equation}
The Bose-Einstein condensation is then also a first-order phase transition in the considered models of gravity if $L\neq0$.

\section{More physical models}
Building upon the discoveries outlined in the preceding section of the paper, one can delve into studying toy-model systems. However, it is crucial to handle the conclusions drawn from such investigations with utmost care, as they have the potential to foster misconceptions. Therefore, overly optimistic bounds on a theory parameter, as discussed in \cite{Wojnar:2024xdy}, or "no-go theorems" for models of gravity with toy-model assumptions \cite{barausse2008curvature,barausse2008no,pani2012surface,Casado-Turrion:2022xkl} should not be decisive in assessing the plausibility of a theory (see e.g. \cite{kim2014physics,olmo2008reexamination,wojnar2023fermi,2020CQGra..37u5002O}).

Therefore, in the following section, we will derive a Fermi EoS with finite temperature corrections and revisit a simple Landau model of liquid helium. This model realistically describes its behavior at low temperatures, in contrast to treating it as an ideal Bose-Einstein condensate.

\subsection{Fermi gas with finite temperature corrections}
In the non-relativistic limit, the energy becomes $E = p^2/2m$ and, consequently, the pressure \eqref{pressure} becomes

\begin{equation}
P= a_1 \left[ \int_0^{\infty} \frac{E^{3/2}}{1+e^{\beta (E-\mu)}} dE + a_2 \int_0^{\infty} \frac{E^{2}}{1+e^{\beta (E-\mu)}}dE \right]\, ,
\end{equation}
where $a_1=\frac{(2m)^{3/2}}{3\pi^2\hbar^3}$ and $a_2=\frac{3 }{4}\sigma d (2m)^{1/2}$. The two integrals above are Fermi integrals with the following solutions:
\begin{align}
\int_0^{\infty} \frac{E^{3/2}}{1+e^{\beta (E-\mu)}} dE =& \frac{2}{5}\mu^{5/2}-\frac{1}{8}\beta^{-1} \mu^{3/2} \ln (1+e^{-\beta \mu})  \\
+&\frac{\pi^2}{4} \beta^{-2} \mu^{1/2} + \frac{3}{4} \beta^{-2} \mu^{1/2} Li_2(-e^{-\beta \mu}) + ... \nonumber\\
\int_0^{\infty} \frac{E^{2}}{1+e^{\beta (E-\mu)}} dE =& \frac{\mu^{3}}{3}
+\frac{\pi^2}{3} \beta^{-2} \mu^{1} + ...
\end{align}

Then, by setting $\mu = E_F$ we can rewrite the pressure as
\begin{eqnarray} \label{pressurefinite}
P_F &=& a_1 \left\{\frac{2 A^{5/2}}{5}\left(\frac{\rho}{\mu_e}\right)^{5/3} \left[1-\frac{5}{16} \psi \ln(1+e^{-1/\psi}) \right. \right. \nonumber \\
&+& \left. \frac{5\pi^2}{8}\psi^2 + \frac{15}{8} \psi^2 Li_2(e^{-1/\psi})
    \right]
\nonumber \\
    &+& \left. \frac{a_2}{3} A^3 \left(\frac{\rho}{\mu_e}\right)^2 (1+\pi^2 \psi^2) \right\} \, ,
\end{eqnarray}
where $A = \frac{(3\pi^2 \hbar^3 N_A)^{2/3}}{2 m_e}$ is a constant and $\psi$ is the degeneracy parameter defined in \eqref{degparameter}. Notice that the first terms of the EoS \eqref{pressurefinite} corresponds to the pressure for $T=0$ in a non-deformed phase space, i.e., the polytrope equation for $n=3/2$, discussed already in \ref{deg}. The first line in the same equation corresponds to the pressure for a finite temperature $T$ in a non-deformed phase space. 

Such an equation of state is used to describe the matter properties in low mass stars and brown dwarfs, in which the finite temperature corrections, providing the time evolution of electron degeneracy, are important, see e.g. \cite{auddy2016analytic,Benito:2021ywe,Kozak:2022hdy}.

\subsection{Liquid helium He \texorpdfstring{$^4$}{TEXT} } \label{hel}

The Landau model \cite{landau2018theory,tisza1947theory} offers a comprehensive microscopic depiction of a two-fluid model near absolute zero. As $T$ approaches zero, the specific heat of liquid helium behaves as $T^3$, characteristic of a phonon gas and experimentally confirmed. However, at finite temperatures, an additional term emerges. Thus, the dispersion relation of quasiparticles as a function of wave number $k$ for $^4$He can be described by:
\[
 \hbar \omega =
  \begin{cases} 
 \hbar ck & \mbox{if } k<< k_0,  \\
 \Delta+ \frac{\hbar^2(k-k_0)^2}{2\gamma}  & \mbox{if } k\approx k_0,
  \end{cases}
\]
where $c$ denotes the sound velocity, while $\Delta$, $k_0$, and $\gamma$ are experimental constants\footnote{We will use the data from \cite{yarnell1959excitations} 
\begin{align*}
    c=239\text{ m s}^{-1},\,\,\rho=144\text{ kg m}^{-3},\,\,\Delta/k_B=8.65\text{K},
    \\ k_0 =
    1.92\times10^{10}\text{ m}^{-1},\,\, \gamma = 1.07\times10^{-27}\text{ kg}.
\end{align*}
}. In the Landau theory, it is posited that the quantum states of $^4$He near the ground state can be treated as those of a non-interacting gas with energy levels given by:
\begin{align}\label{uint}
U=E_0+\sum_k \hbar \omega_k \langle n_k \rangle 
= E_0 + \frac{V}{2\pi^2} \int^\infty_0 \frac{k^2\hbar\omega_k}{e^{\beta\hbar\omega_k}-1}\frac{dk}{(1-\sigma\hbar k)}.
\end{align}
Here, $ \hbar \omega_k $ represents the elementary excitation energy with wave vector $k$ and occupation number $\langle n_k \rangle$. In the second equality, the deformation of the phase space has been taken into account. Now, let us compute the internal energy and its gravity corrections at low temperatures. In this context, only contributions from the phonon and roton components \cite{cohen1957theory,yarnell1959excitations} affect the energy in Eq. \eqref{uint}. They are given, respectively, as:
\begin{align}
     E_\text{phonon}&=\frac{V}{2\pi^2}\left( \frac{\pi^4(k_BT)^4}{15\hbar^3 c^3} + 24\sigma \frac{(k_BT)^5\zeta(5)}{c^4\hbar^3}\right),\\
       \frac{E_\text{roton}}{V}&\approx \frac{k_0^2\Delta}{\pi}\sqrt{\frac{\gamma k_BT}{2\pi\hbar^2}}e^{-\frac{\Delta}{k_BT}}(1+\sigma \hbar k_0).
\end{align}
Using the expression \eqref{therm4}, we derive the specific heat for liquid helium in low temperature
\begin{align}\label{sheat}
    C_{\text{He}^4}&=20.7 T^3+ \frac{387\times10^{3}}{T^{3/2}} e^{-8.85/T} 
    +\sigma (5.73\times10^{-24}T^4 \nonumber \\&+\frac{7.83\times10^{-19}}{T^{3/2}} e^{-8.85/T}). 
\end{align}

%%%%%%%%%%%%%%%%%%%%%%%%%%%%%%%%%%%%%%%%%%
\section{Discussion and conclusions}
In this paper, our focus was to investigate the impact of Ricci-based gravity theories, such as Palatini $f(R)$ and Eddington-inspired Born-Infeld models, in conjunction with linear Generalized Uncertainty Principle (GUP) models, on systems governed by Bose and Fermi statistics. By leveraging the recently established correspondence between modified gravity and GUP models, we developed a formalism to analyze ideal Bose and Fermi gases. As anticipated, the inclusion of modified gravity or linear GUP resulted in the introduction of additional terms to the familiar expressions. We outlined these modifications for general Fermi gases, degenerate Fermi gases, Boltzmann gases, as well as Bose gases including phonons, photons, Bose-Einstein condensates, and liquid helium.

It's worth noting that the thermodynamic framework derived from this correspondence is best suited for non-relativistic systems or those characterized by low curvature regimes, such as planets, brown dwarfs, active stars, and white dwarfs. Even white dwarfs, despite their dense nature, can still be considered within a non-relativistic regime due to their size. Consequently, the compactness criterion applicable to our formalism is $\mathcal C<<1$.

This intriguing connection between Modified Gravity and GUP models opens up avenues for testing gravitational proposals through various tabletop experiments, some of which have already been explored and documented in the literature. Ongoing research in this direction aims to delve deeper into the implications of Modified Gravity on the microscopic properties of matter, seeking further validation and insights.

\section*{Acknowledgements}
 AW acknowledges financial support from MICINN (Spain) {\it Ayuda Juan de la Cierva - incorporaci\'on} 2020 No. IJC2020-044751-I.

\bibliographystyle{apsrev4-1}

\end{document}